\documentclass{article}
\usepackage{ijcai21}

\usepackage{times}
\usepackage{soul}
\usepackage{url}
\usepackage[hidelinks]{hyperref}
\usepackage[utf8]{inputenc}
\usepackage{subfigure}
\usepackage{amsmath}
\usepackage{multirow}
\usepackage[small]{caption}
\usepackage{graphicx}
\usepackage{amsmath}
\usepackage{amsthm}
\usepackage{booktabs}
\usepackage{algorithm}
\usepackage{algorithmic}
\usepackage{color, xcolor}
\usepackage{amssymb} 
\colorlet{dark-blue}{blue!50!black}

\title{Location Predicts You: \\
Location Prediction via Bi-direction Speculation and Dual-level Association}

\author{
Xixi Li$^{1}$
\and
Ruimin Hu$^{1}$\thanks{Corresponding Author}\and
Zheng Wang$^{2,3}$\And
Toshihiko Yamasaki$^{2,3}$
\affiliations
$^1$National Engineering Research Center for Multimedia Software (NERCMS), \\School of Computer Science, Wuhan University\\
${^2}$Research Institute for an Inclusive Society through Engineering (RIISE), The University of Tokyo \\ 
${^3}$Department of Information and Communication Engineering, The University of Tokyo \\
\emails
\{xixil, hrm\}@whu.edu.cn,
wangzwhu@gmail.com,
yamasaki@cvm.t.u-tokyo.ac.jp
}

\begin{document}

\maketitle

\begin{abstract}
Location prediction is of great importance in location-based applications for the construction of the smart city. To our knowledge, existing models for location prediction focus on users' preferences on POIs from the perspective of the human side. However, modeling users' interests from the historical trajectory is still limited by the data sparsity. Additionally, most of existing methods predict the next location according to the individual data independently, but the data sparsity makes it difficult to mine explicit mobility patterns or capture the casual behavior for each user. To address the issues above, we propose a novel Bi-direction Speculation and Dual-level Association method (BSDA), which considers both users' interests in POIs and POIs' appeal to users. Furthermore, we develop the cross-user and cross-POI association to alleviate the data sparsity by similar users and POIs to enrich the candidates. Experimental results on two public datasets demonstrate that BSDA achieves significant improvements over state-of-the-art methods.
 \end{abstract}

\section{Introduction}
Human mobility modeling on spatiotemporal data has gained significant research interests due to its importance for human behavior understanding and applications for the construction of smart city~\cite{yu2019adaptive,wang2020beyond,wang2021very}. One of the most important applications is the next location prediction, which aims to predict the next Point-of-Interest (POI) someone tends to visit based on his/her historical trajectory data.

In recent years, plenty of research efforts have been made on the location prediction~\cite{yang2020location}, including Matrix Factorization based~\cite{ren2017context}, Markov Chain (MC) based~\cite{zhang2014lore,chen2014nlpmm} and neural network based methods~\cite{karatzoglou2018seq2seq}. To our knowledge, these methods only investigated how the user chooses places, \textit{i.e.}, focusing on the users' preferences on POIs \textit{from the perspective of the human side}. Unfortunately, despite the success of these works for enriching the representation, modeling users' interests from the human trajectory history is still limited by the data sparsity due to the fact that most users share few records online. In addition, most of methods predicted someone's next location according to his/her own trajectory independently. But the data sparsity makes it difficult to mine explicit mobility patterns for each user or to capture the casual behavior of the user. Hence, the next location prediction is still a challenging task.

To address the issues above, in this paper, we propose a novel Bi-direction Speculation and Dual-level Association method (BSDA). The characteristics of our method are reflected in the following aspects: 1) \textit{From the Perspective of the Location Side}: we consider that the POI appeals or chooses the users autonomously. If we set a POI as the subject, another model should be capable of predicting the next user who will visit this POI, according to the historical data of this POI. 2) \textit{Cross-User Association}: we believe that some users have similar interests and they may visit the same places. When we can not predict one's next location accurately due to his incomplete historical trajectory data, the POI candidate of users who are similar to him might be his choice. 3) \textit{Cross-POI Association}: it is also reasonable that similar places appeal to common users. 
Hence, the most similar POIs also benefit the prediction of next coming users.

\tabcolsep=2pt
\begin{figure*}[t]
    \centering
    \begin{tabular}{cccc}
		\includegraphics[width=0.22\textwidth]{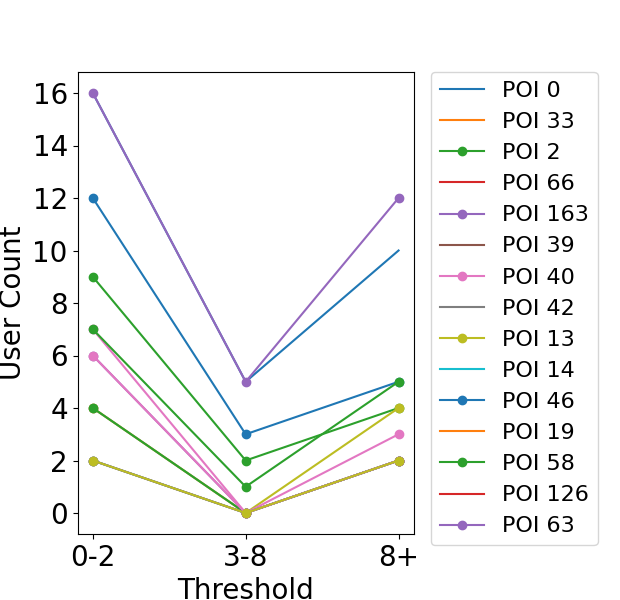}&
		\includegraphics[width=0.26\textwidth]{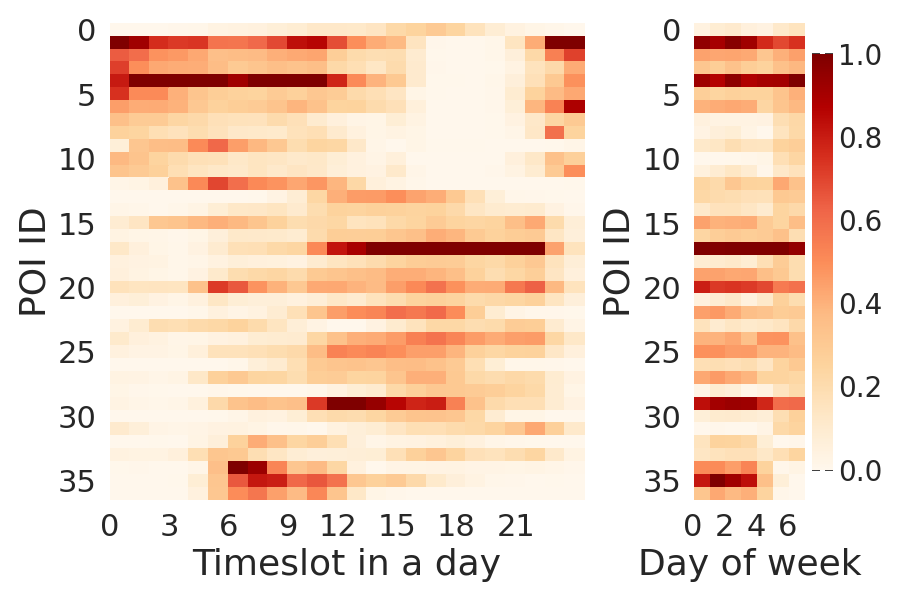} &
        \includegraphics[width=0.25\textwidth]{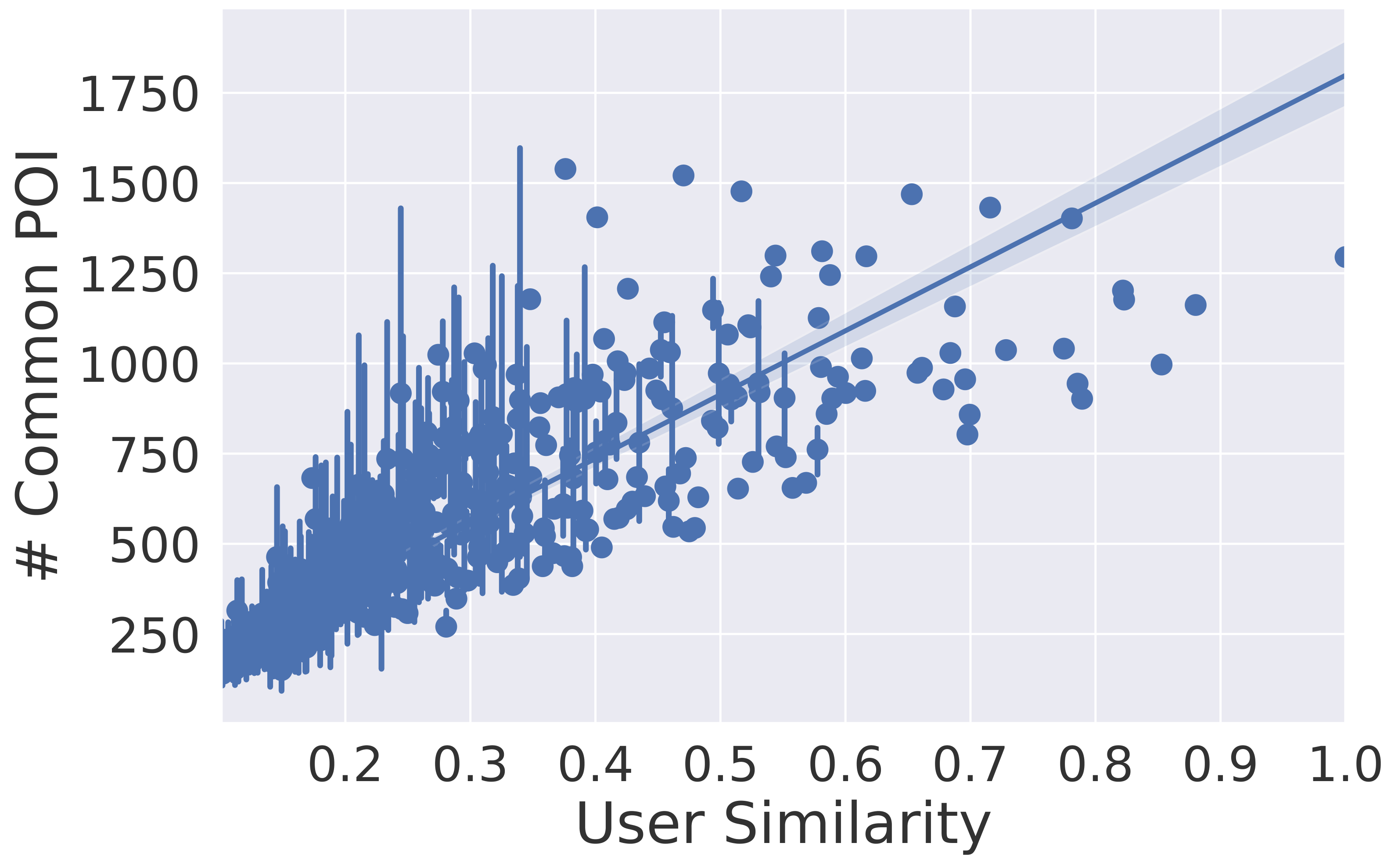} &
        \includegraphics[width=0.24\textwidth]{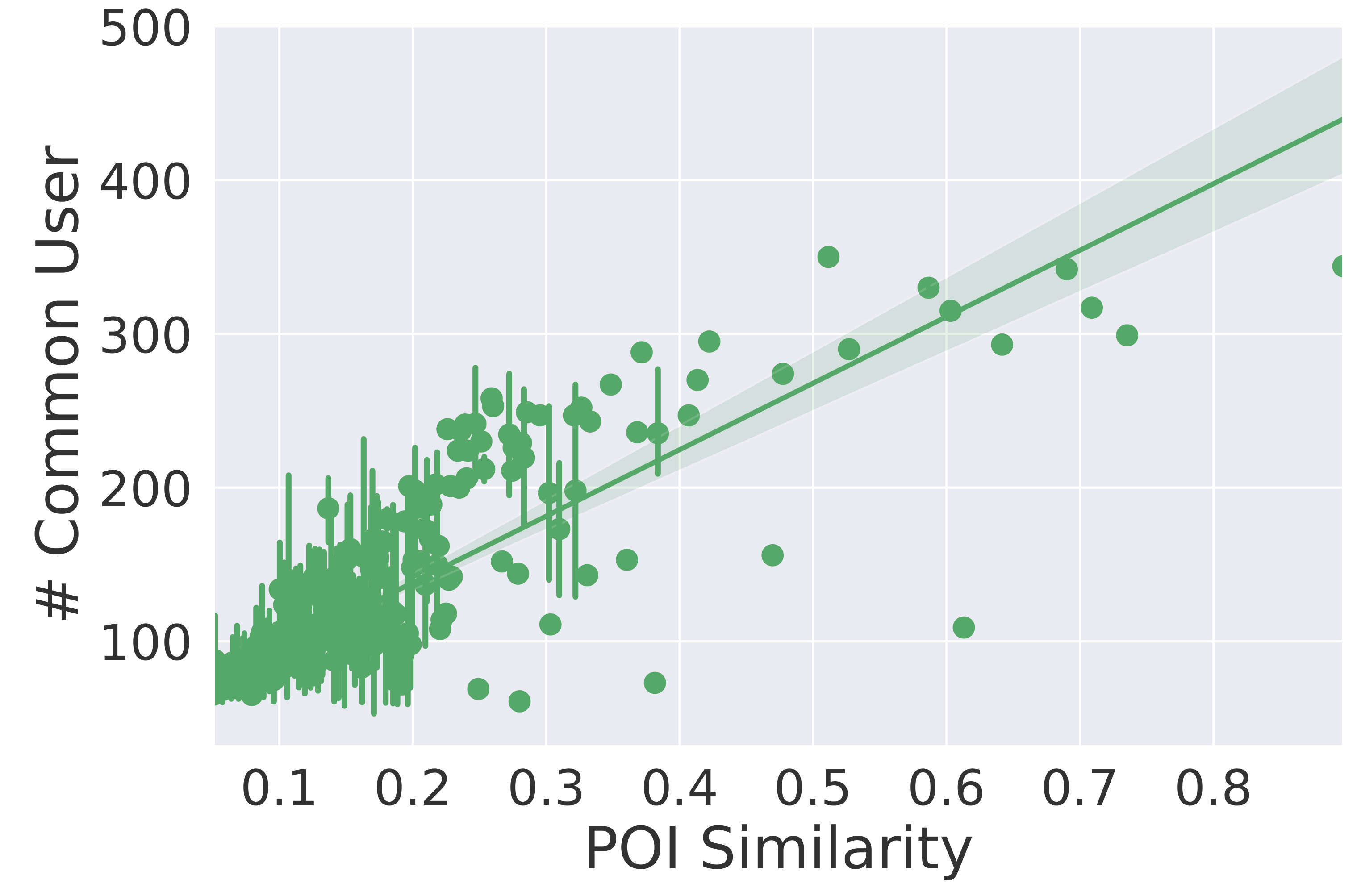}\\
        \small{(a) User Visiting Times} & 
        \small{(b) POI's Temporal Pattern}  & 
        \small{(c) User-sim \textit{vs.} \#Comm. POI} &
        \small{(d) POI-sim \textit{vs.} \#Comm. User} \\   
    \end{tabular}
    \caption{\label{fig:motivation}The statistical data to show the motivation of our method.}
\end{figure*}

\subsubsection{Motivation}
We conduct some preliminary statistic analysis on the Gowalla dataset~\cite{cho2011friendship} to demonstrate our motivations described above.

Firstly, we investigate how the POI appeals or chooses the visiting users. We count the visiting times of the users for several randomly selected POIs. Figure~\ref{fig:motivation}(a) shows the results of 15 POIs. The horizontal axis stands for the threshold of the visiting times of each user, while the vertical axis denotes the number of users visiting the corresponding POI. The figure demonstrates that for a certain POI, some users visit frequently while the others visit occasionally, and only a few users visit 3-8 times. Here is an example to illustrate this phenomenon. A library appeals to people who like reading but does not attract those who don't have relevant habits. We also explore temporal patterns in POIs' behavior by calculating the densities of visitors at different time slots in a day and on different days in a week. Figure~\ref{fig:motivation}(b) shows the statistic results of randomly selected 35 POIs. The figure demonstrates that POIs have different temporal patterns as their distinct difference in densities on the two time scales. It is easy to understand that an amusement park often has more visitors at weekends than on weekdays, while an office building always has an opposite pattern. Additionally, a bar is usually busy at night, while a museum only accommodates visitors in the morning. These findings show that POI has its behavior patterns to appeal to a certain kind of users. Intuitively, it is reasonable that the task can be considered from the perspective of the location side.

Secondly, we intend to validate that similar users may visit the same places. We judge the similarity of each pair of users according to whether they visit the same place on the same day. The more common places they visit, the more similar the users' visiting patterns are. We also count the number of common POIs of each pair of users. Figure~\ref{fig:motivation}(c) shows the results of user similarity \textit{vs.} \#common POI. Figure~\ref{fig:motivation}(c) easily demonstrates that the user similarity and the number of common POIs are proportional to each other (highly related). Hence, the next POIs of one's most similar users can be considered as the choice, on condition that we can not predict his/her next POI accurately due to the incomplete historical trajectory data.

Thirdly, we evaluate whether similar places may appeal to common users. We judge the similarity of each pair of POIs according to whether they appeal to the common users on the same day. The more common visitors they have, the more similar their behavior patterns are. We also count the number of common users of each pair of POIs. Figure~\ref{fig:motivation}(d) shows the results of POI similarity \textit{vs.} \#common user. Figure~\ref{fig:motivation}(d) easily demonstrates that the POI similarity and the number of common users are proportional to each other (highly related). Hence, the next visitors of similar POIs can be considered as the choice of the focused POI.

To this end, we propose a novel Bi-direction Speculation and Dual-level Association method (BSDA) for the next location prediction by mining both users' interests in POIs and POIs' appeal to users.
From the point of the human side, we pay attention to when and where the user is likely to go.
While from the point of the location side, we focus on when and who will visit this place.
Specifically, we first develop two networks to explore users' interests and POIs' appeal separately.
In addition, we exploit the cross-user and cross-POI association to alleviate the data sparsity by mining similar users and POIs to enrich the candidates. In particular, we learn the POI similarity and user similarity matrices through the historical trajectory data and fuse them with the outputs of two networks to predict the next location of all related users. We verify the effectiveness of our proposed model on two public check-in datasets and experimental results show that BSDA outperforms state-of-the-art methods for the location prediction task.

The contributions of our paper are summarized as follows: 1) We first investigate the phenomena of POIs' behavior patterns about appeal to users, user similarity and POI similarity. 2) We propose a novel Bi-direction Speculation and Dual-level Association method (BSDA) for location prediction by mining both users' interests and POIs' appeal, as well as fusing with cross-user and cross-POI association. 3) Evaluations on two public real-world datasets show that BSDA achieves significant improvements over the state-of-the-art methods.

\begin{figure*}[t]
	\centering
	\footnotesize{
	\begin{tabular}{c}
		\includegraphics[width=0.96\textwidth]{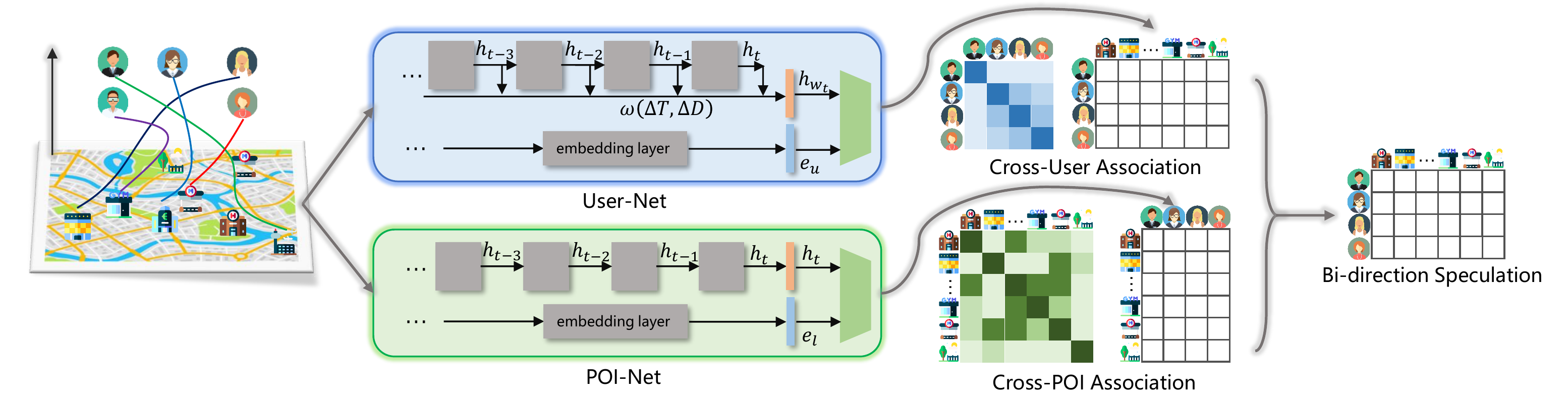} \\
	\end{tabular}}
	\caption{The framework of our Bi-direction Speculation and Dual-level Association method (BSDA). It consists of the User-Net and the POI-Net to speculate the next POI where a user will go or the next user who will visit the POI, respectively. It also contains the cross-user and cross-POI association parts to adjust the results. Finally, bi-direction speculations are fused for the location prediction.}
	\label{fig: model} 
\end{figure*}

\section{Related Work}
Current methods on location prediction can be mainly categorized into three groups: Matrix Factorization (MF) based, Markov Chain (MC) based, and RNN based. MF-based methods~\cite{koren2009matrix,liu2013learning} are good at capturing users' general preferences but cannot mine users' sequential patterns. MC-based methods~\cite{zhang2014lore,chen2014nlpmm,he2016fusing} are capable of capturing the sequential transitions but the first-order MCs fail to learn the long-term preference. By combining the power of MF and MC, \cite{rendle2010factorizing} proposed a factorizing personalized MCs to obtain both long-term preference and short-term transitions. \cite{wang2015learning} made an extension with global user vectors to learn users' hierarchical representation. \cite{feng2015personalized} also proposed a personalized ranking metric embedding for personalized recommendation by integrating sequential information, individual preference and geographical influence. However, these methods mainly focus on modeling the user-POI interaction or the transitions between successive POIs to mine users' preferences, limited by the data sparsity since most users share few records online.

RNN~\cite{zhang2014sequential} and its variants LSTM~\cite{hochreiter1997long} and GRU~\cite{cho2011friendship}, have shown great advantages in sequential modeling task. For instance, \cite{yao2017serm} learned temporal regularity, activity semantics, and user preference by embedding sequential context and temporal information together into LSTM.
\cite{sun2020go} incorporated both long- and short-term preferences and considered  geographical relations between two nonconsecutive POIs.
Other deep models such as memory network and attention mechanism~\cite{vaswani2017attention} were also widely employed~\cite{ying2018sequential}. \cite{huang2018csan} proposed to discriminate the significance of individual user behaviors using the feature-wise masked self-attention.
Similarly, \cite{yu2019adaptive} applied an attention-based adaptive fusion to dynamically combine users' long- and short-term interests. These methods mainly pay attention to modeling users' behavior patterns by mining the latent representation of users' interests.
Some methods leverage contextual information to enrich the representation \cite{yu2020category}.
\cite{lim2020stp} proposed a graph attention network that not only exploits personalized user preferences but also explores new POIs from other users.
However, the problem is studied only from the perspective of the human side, still limited by the data sparsity.

\cite{feng2017poi2vec} proposed POI2Vec which can predict the potential future visitors for a location by modeling the POI sequential transition. Although POI2Vec introduces the task of predicting future visitors, its one-direction single module for modeling users' patterns is more close to our User-net. However, we design two parallel modules to treat users and POIs as subjects to do the bi-direction speculation.
Furthermore, three points about our dual-level association need to be emphasized. 1) Our proposed association is to adjust the POI or user candidate list, which occurs after training while sharing parameters or using embedding is during training. 2) In POI-net and User-net, our proposed similarities is to provide more possible candidates based on patterns, not just shared data. 3) Our dual-level association is time-aware, \textit{i.e.}, the output candidate list is dynamically adjusted according to the similar behavior at a given time.

To sum up, our method has the following advantages. To our best knowledge, it is the first time that the location prediction problem is considered from both the perspective of the human side and the location side. We propose the bi-direction speculation, \textit{i.e.}, mining both users' interests in POIs and POIs' appeal to users.
Moreover, we exploit the cross-user and cross-POI association to alleviate the data sparsity by mining similar users and POIs to enrich the candidates.

\section{Proposed Methodology}
\subsection{Problem Formulation}
Let $\mathcal{U} = \left\{ {{u_1},{u_2},\cdots,{u_M}} \right\}$ be the set of $M$ users and
$\mathcal{L} = \left\{ {{l_1},{l_2},\cdots,{l_N}} \right\}$ be the set of $N$ POIs.

\paragraph{Check-in point.}
When a user $u$ visits a POI $l$ at time $t$, the check-in point is represented as $r_{t}^{u}=\left \langle u, t, lon, lat, l \right \rangle$, where $lat$ and $lon$ represent the geographical coordinates of $l$.

\paragraph{Trajectory.}
Given a user $u$, the trajectory is denoted as a set of chronologically ordered check-in points $\mathcal{R}_{u}=\left \{ r_{t_{1}}^{u}, r_{t_{2}}^{u}, \cdots , r_{t_{T}}^{u} \right \}$, where $T$ is the length of trajectory.

\paragraph{Problem definition.}
Formally, given a user's trajectory history in the last $T-1$ steps, the primary goal of the next location prediction task is to predict the top-k most likely POIs $\{L_{k}\}$ that this user $u$ will visit at the next time $t_{T}$.

\subsection{Framework Overview}
We employ a novel Bi-direction Speculation and Dual-level Association method (BSDA) for the next location prediction. The overall framework is depicted in Figure~\ref{fig: model}. It mainly consists of three parts: 1) the User-Net and cross-user association for modeling users’ interests to speculate where a user is likely to go, 2) the POI-Net and cross-POI association for modeling POIs’ appeals to speculate who will visit the POI, and 3) the fusion layer for integrating the results of bi-direction speculations.

\subsection{User-Net and Cross-User Association}
The User-Net is designed for exploring users' interests in POIs by modeling users' spatiotemporal behavior patterns, which belongs to the traditional location prediction task. Its input is the trajectory history of users and the output is the POI candidate score list. Motivated by the success of Flashback \cite{yang2020location}, our User-Net uses RNNs to capture the sequential patterns and compute the weighted average of the historical hidden states as the aggregated output $h_{w_{t}}$ with a decay weight $\omega \left ( \Delta T, \Delta D \right )$ to predict the next location.
The decay weight is designed to measure the predictive power of the hidden state $h_{j}$ to the current one $h_{t}, j < t$:
 $\omega \left ( \Delta T_{i, j}, \Delta D_{i, j} \right )=\operatorname{hvc}\left ( 2\pi \Delta T_{i, j} \right )e^{-\alpha \Delta T_{i, j}}e^{-\beta \Delta D_{i, j}}$,
where $\Delta T$ denotes the temporal interval, $\Delta D$ denotes the geographical distance. The Havercosine term $\operatorname{hvc}\left ( \cdot \right )$ is to capture the periodicity of user behavior. The exponential terms are to capture the impact of historical check-ins. Therefore, the User-Net is capable of capturing the sequential patterns, long-term temporal dependencies and the impact of historical check-ins on the next location prediction.

For each user, we first split the historical trajectory $\mathcal{R}_{u}$ into sub-sequences as the input of the RNN units. At each step $t$, the hidden state is denoted as $h_{t}$.
Then we leverage the spatiotemporal contexts to generate the decay weight $\omega \left ( \Delta T, \Delta D \right )$ for calculating the aggregated output $h_{w_{t}}=\frac{\sum_{j=0}^{t}w_{j}\ast h_{j}}{\sum_{j=0}^{t}w_{j}}$. Furthermore, to model personalized preference of each user, we concatenate the user embedding $\boldsymbol{e_{u}}$ with the aggregated outputs $h_{w_{t}}$ and then feed it into a fully connected layer to generate the POI candidate score list $\mathbf{s}_{U}^{u}$ for user $u$.

Nevertheless, due to the data sparsity, the prediction performance will be unsatisfactory when a user has few check-ins.
To tackle this problem, we propose the cross-user association to provide more possible POI candidates according to the most similar users.
The user similarity is defined according to whether the user pair visit the same place on the same day. 
Specifically, the similarity between user $u_{m}$ and user $u_{n}$ is: $\tau _{m,n}=\frac{\left | L_{m}\cap L_{n} \right |}{\left | L_{m} \right |}$, where $L_{m}$ denotes the set of all locations that user $m$ visits.
We first generate the user similarity matrix $\boldsymbol{Corr}_{U}$ for measuring how similar the user pair's interests in POIs are, where $\boldsymbol{Corr}_{U}\in \mathbb{R}^{M\times M}$, each item $\tau _{m,n}, m,n\leq M$ in $\boldsymbol{Corr}_{U}$ denotes the similarity between a user pair.
$\boldsymbol{Corr}_{U}$ is normalized to $\left [ 0,1 \right ]$.

On the other hand, we can generate a POI candidate score matrix $\mathbf{S}_{U}$, where $\mathbf{S}_{U}=\left [ \mathbf{s}_{U}^{u_{1}};\mathbf{s}_{U}^{u_{2}};\cdots ;\mathbf{s}_{U}^{u_{M}} \right ]$, $\mathbf{S}_{U}\in \mathbb{R}^{M\times N}$, $\mathbf{s}_{U}^{u}$ denotes the POI candidate score list of user $u$ at time $t_{T}$, $u\in \mathcal{U}$.
We believe that users with similar interests in POIs are likely to visit the same places.
If we can not predict one's next location accurately due to his incomplete trajectory history, the next POI candidates of his most similar users can be considered as the choice.
Therefore, we multiply the POI candidate score matrix $\mathbf{S}_{U}$ with the user similarity matrix $\boldsymbol{Corr}_{U}$ to get the adjusted score matrix $\hat{\mathbf{S}}_{U}$:
\begin{equation}
    \hat{\mathbf{S}}_{U} = \boldsymbol{Corr}_{U}\ast \mathbf{S}_{U}
\end{equation}
where $\hat{\mathbf{S}}_{U}\in \mathbb{R}^{M\times N}$. Finally, for a user $u$, we can get the adjusted POI candidate score list $\hat{\mathbf{s}}_{U}^{u}$ from $\hat{\mathbf{S}}_{U}$.

\subsection{POI-Net and Cross-POI Association}
Based on the observations on the POIs in Figure~\ref{fig:motivation}(a) and Figure~\ref{fig:motivation}(b), we derive the assumption that a POI can appeal or choose the users autonomously and there exist patterns in POIs’ behavior. 
Hence, we intend to explore POIs' appeals and predict the next user who will visit the POI. 
From the perspective of the location side, we assume that POIs’ appeals are dynamically changed over time. Therefore, we exploit POIs’ dynamic appeals by using RNNs to capture temporal patterns considering different time slots in a day and different days in a week.

Given the historical data, we first represent POIs' records in form of user sequences. For each POI $l$, we denote its visiting point as $a_{t}^{l}=\left \langle l, u, s, d \right \rangle$. Here we represent time $t$ on two scales, \textit{i.e.} $s$ denotes the time slot in a day and $d$ denotes the day in a week.
The whole visiting history of POI $l$ is denoted as a set of chronologically ordered visiting points $A_{l}=\left \{ a_{t_{1}}^{l}, a_{t_{2}}^{l}, \cdots , a_{t_{{T}'}}^{l} \right \}$, where ${T}'$ is the length of visiting history.
In the POI-Net, given a POI’s visiting history in the last ${T}'-1$ steps, the primary goal of the POI-Net is to predict the top-k most likely users $U_{k}$ who may visit this POI at the next time $t_{{T}'}$. In this paper, we take the basic RNN to model POIs' behavior patterns and use the current hidden state $h_{t}$ to predict the next user.
Besides, to model the personalized appeal of each POI, we concatenate the POI embedding $\boldsymbol{e_{l}}$ with the current hidden state $h_{t}$ and then feed it into a fully connected layer to generate the user candidate score list $\mathbf{s}_{L}^{l}$ for POI $l$.

However, the performance of the user prediction is unsatisfactory because some POIs are checked in very few times.
To deal with this problem, after capturing the POIs' appeal by the RNNs, we further employ a cross-POI association approach to improve the performance of the next user prediction.
The POI similarity is according to whether the POI pair are visited by the same user on the same day.
Specifically, the similarity $\tau _{m,n}$ between POI $l_{m}$ and POI $l_{n}$ is the number of days on which POI $m$ and $n$ appeal the same user, \textit{i.e.}, $U_{m,d} \cap U_{n,d} \neq \varnothing$, 
where $U_{m,d}$ denotes the set of visitors of location $m$ on day $d$.
We first generate the POI similarity matrix $\boldsymbol{Corr}_{L}$ for measuring how similar the POI pair's appeal to users are, where $\boldsymbol{Corr}_{L}\in \mathbb{R}^{N\times N}$, each item $\tau _{m,n}, m, n\leq N$ in $\boldsymbol{Corr}_{L}$ denotes the similarity between a POI pair. $\boldsymbol{Corr}_{L}$ is normalized to $\left [ 0,1 \right ]$.
Meanwhile, we generate a user candidate score matrix $\mathbf{S}_{L}$ which consists of the user candidate score list of each POI in set $\mathcal{L}$, \textit{i.e.} $\mathbf{S}_{L}=\left [ \mathbf{s}_{L}^{l_{1}};\mathbf{s}_{L}^{l_{2}};\cdots ;\mathbf{s}_{L}^{l_{N}} \right ]$, $\boldsymbol{S}_{L}\in \mathbb{R}^{N\times M}$, $\mathbf{s}_{L}^{l}$ denotes the user candidate score list of POI $l$ at time $t_{{T}'}$, $l\in \mathcal{L}$.
Under the assumption that similar POIs appeal to common users, we can provide more possible user candidates for a POI $l$ by using the next user candidates of its most similar POIs.
Hence, we multiply the POI similarity matrix $\boldsymbol{Corr}_{L}$ and the user candidate score matrix $\mathbf{S}_{L}$ to get the adjusted score matrix $\hat{\mathbf{S}}_{L}^{l}$:
\begin{equation}
    \hat{\mathbf{S}}_{L} = \boldsymbol{Corr}_{L}\ast \mathbf{S}_{L}
\end{equation}
where $\hat{\mathbf{S}}_{L}\in \mathbb{R}^{N\times M}$.

\begin{table*}
\begin{center}
\small
\begin{tabular}{c|c|c|cccc|cccc}
\hline
\multirow{2}{*}{Methods} & \multirow{2}{*}{Category}& \multirow{2}{*}{Reference} & \multicolumn{4}{c}{Gowalla}	& \multicolumn{4}{|c}{Foursquare}\\
\cline{4-11}
  & & & Acc@1 & Acc@5 & Acc@10 &MRR & Acc@1 & Acc@5 & Acc@10 &MRR\\
\hline
\hline
BPR~\cite{rendle2009bpr} & MF &UAI 2009 &0.0131 &0.0363 &0.0539 &0.0235 &0.0315 &0.0828 &0.1143 &0.0538 \\
FPMC~\cite{rendle2010factorizing}& MF+MC & WWW 2010 &0.0479  &0.1668   &0.2411   &0.1126   &0.0753   &0.2384   &0.3348   &0.1578   \\
RNN~\cite{zhang2014sequential} & RNN &AAAI2014 &0.0881 &0.2140 &0.2717 &0.1507 &0.1824 &0.4334 &0.5237 &0.2984 \\
PRME~\cite{feng2015personalized} &MC+ME & IJCAI 2015 &0.0740  &0.2146   &0.2899   &0.1503   &0.0982   &0.3167   &0.4064   &0.2040   \\
STRNN~\cite{liu2016predicting} &RNN & AAAI 2016 &0.0900  &0.2120  &0.2730  &0.1508  &0.2290  &0.4310  &0.5050  &0.3248  \\
DeepMove~\cite{feng2018deepmove} &RNN & WWW 2018 &0.0625  &0.1304  &0.1594  &0.0982  &0.2400  &0.4319  &0.4742  &0.3270  \\
LBSN2Vec~\cite{yang2019revisiting} & graph & WWW 2019 &0.0864  &0.1186   &0.1390   &0.1032   &0.2190   &0.3955   &0.4621   &0.2781   \\
STGN~\cite{zhao2020go} & RNN & TKDE 2020 &0.0624  &0.1586  &0.2104  &0.1125  &0.2094  &0.4734  &0.5470  &0.3283  \\
Flashback~\cite{yang2020location} & RNN & IJCAI 2020 &0.1158 &0.2754 &0.3479 &0.1925 &0.2496 &0.5399 &0.6236 &0.3805 \\
\hline
BSDA &RNN  & Ours &\textbf{0.1454} &\textbf{0.3531} &\textbf{0.4192} &\textbf{0.2413} &\textbf{0.3068} &\textbf{0.6612} &\textbf{0.7136} &\textbf{0.4505}\\
\hline
\end{tabular}
\end{center}
\caption{Comparisons on two datasets (evaluated by Acc@1, Acc@5, Acc@10, and MRR).}
\label{table: comparison of baselines}
\end{table*}

\subsection{Bi-direction Speculation}
After predicting the next POI that a user will visit from the User-Net and cross-user association, and the next user that a POI will appeal from the POI-Net and cross-POI association, we integrate them for bi-direction speculation to improve the performance of the next location prediction.
In this way, we achieve a more reasonable and precise location prediction result.
In detail, after generating the adjusted POI candidate score matrix $\hat{\mathbf{S}}_{U}$ 
and the user candidate score list $\hat{\mathbf{S}}_{L}$, we apply a fusion layer to integrate the two by maxpooling.
The maxpooling operation is to take the lager value for the corresponding position in $\hat{\mathbf{S}}_{U}$ and ${\hat{\mathbf{S}}_{L}}^\top$,
where the output ${\mathbf{S}}_{F}\in \mathbb{R}^{M\times N}$.
The key formula of the hybrid representation $\mathbf{S}_{F}$ is as follows:
\begin{equation}
    \mathbf{S}_{F} = \operatorname{maxpooling}(\hat{\mathbf{S}}_{U} + {\hat{\mathbf{S}}_{L}}^\top)
\end{equation}
For a user $u$, we can get the final hybrid POI candidate score list $\mathbf{s}_{F}^{u}$ from $\mathbf{S}_{F}$.
Finally, we rank possible POIs in $\mathbf{s}_{F}^{u}$ according to the scores and get the top-k most likely POIs $\left \{ L_{k} \right \}$ for location prediction.

\section{Experiments and Analysis}
\subsection{Experimental Settings}
\paragraph{Datasets.}
We conduct experiments on two widely-used real-world check-in datasets: \textbf{Gowalla} and \textbf{Foursquare}. Gowalla \cite{cho2011friendship} is collected from February 2009 to October 2010 from the Gowalla Website. Foursquare \cite{yang2014modeling} is collected from April 2012 to January 2014. They both contain check-in records in form of userID, POIID, latitude, longitude, and timestamp. Following the setting in \cite{yang2020location}, we eliminate inactive users who have records less than 100. We take the first 80\% check-ins as the training set, the other 20\% as the testing set.

\paragraph{Evaluation Metrics and Parameters.}
We evaluate the effectiveness of the proposed method by using average Accuracy@K and Mean Reciprocal Rank (MRR)~\cite{yang2020location}. 
$Acc@K$ is the percentage of correct predictions for a list of predictions with length $K$: $Acc@K=\frac{N_{correct}}{K}$.
To get more convincing results, we repeat each experiment 15 times and take the average metric values for evaluations. We empirically set the dimension of hidden states for RNN units and the dimension of embeddings as 10. Parameters in flashback block including temporal decay $\alpha$ and spatial decay $\beta$ follows the setting in \cite{yang2020location}.

\paragraph{Compared Methods.}
We compare our proposed BSDA with the following methods: 1) BPR \cite{rendle2009bpr} is a popular matrix factorization method. 
2) FPMC \cite{rendle2010factorizing} is a personalized Markov Chain algorithm by combining the matrix factorization and Markov chain.
3) RNN \cite{zhang2014sequential} is a vanilla RNN architecture. 
4) PRME \cite{feng2015personalized} is a personalized ranking metric embedding method for modeling the sequential transitions. 
5) STRNN \cite{liu2016predicting} extends RNN and models local temporal and spatial contexts. 
6) DeepMove \cite{feng2018deepmove} is an attentional recurrent network for mobility prediction. 
7) LBSN2Vec \cite{yang2019revisiting} is a hypergraph embedding approach combining the random-walk-with-stay scheme and node embedding learning for feature learning.
8) STGN \cite{zhao2020go} is a spatiotemporal gated network by enhancing LSTM. 
9) Flashback \cite{yang2020location} is an RNN-based model for modeling sparse user mobility traces by doing flashbacks on hidden states in RNNs.

\begin{table*}
\begin{center}
\small
\begin{tabular}{c|cccc|cccc}
\hline
\multirow{2}{*}{Methods}  & \multicolumn{4}{c}{Gowalla}	& \multicolumn{4}{|c}{Foursquare}\\
\cline{2-9}
~  &  Acc@1 & Acc@5 & Acc@10 &MRR & Acc@1 & Acc@5 & Acc@10 &MRR\\
\hline
BSDA \textit{w/o} Cross-POI Association   &0.1279 &0.3287 &0.3994 &0.2212 &0.2602 &0.5842 &0.6814 &0.4207 \\
BSDA \textit{w/o} Cross-User Association  &0.1257 &0.3315 &0.3857 &0.2207 &0.2570 &0.5787 &0.6835 &0.4230 \\
BSDA \textit{w/o} user prediction  &0.1215 &0.2919 &0.3721 &0.2001 &0.2545 &0.5506 &0.6329 &0.3881 \\
\hline
BSDA & \textbf{0.1454} &\textbf{0.3531} &\textbf{0.4192} &\textbf{0.2413} &\textbf{0.3068} &\textbf{0.6612} &\textbf{0.7136} &\textbf{0.4505}\\
\hline
\end{tabular}
\end{center}
\caption{Results of variants on two datasets (evaluated by Acc@1, Acc@5, Acc@10, and MRR)}
\label{table: result of variants}
\end{table*}

\subsection{Performance Comparison}
The results of our proposed model BSDA and compared methods on two datasets evaluated by Accuracy@K and MRR are shown in Table~\ref{table: comparison of baselines}. From the table we can observe that: 1) All RNN-based methods including RNN, STRNN, DeepMove, STGN and Flashback perform better than FPMC since they can learn high-order sequential transitions from users' historical trajectories while FPMC can only learn low-order transitions according to the latest check-in. 2) Methods obtain better performances on Foursquare.  This is probably because the data in Gowalla is more sparse. 3) BSDA outperforms all other methods on both datasets, showing its effectiveness for the next location prediction. It improves 25.56\%, 28.21\%, 20.49\%, 22.91\%, at Acc@1, Acc@5, Acc@10, MRR respectively compared with Flashback on Gowalla dataset and improves 22.92\%, 22.46\%, 14.43\%, 18.40\%, at Acc@1, Acc@5, Acc@10, MRR respectively on Foursquare dataset. The significant improvements of BSDA indicate that it can capture more useful information via bi-direction speculation and dual-level association.

\subsection{Model Analysis and Discussion}
\subsubsection{Effectiveness of Key Components}
We further implement three variants to evaluate the contribution of each key component in our proposed BSDA to the performance, including the cross-POI association, the cross-user association and the next user prediction.
The variants are as follows: 1) BSDA w/o Cross-POI Association: we remove the cross-POI association, integrating the User-Net, the cross-user association, and the POI-Net for prediction. 2) BSDA w/o Cross-User Association: we remove the cross-user association, integrating the User-Net, the POI-Net, and the cross-POI association for prediction. 3) BSDA w/o user prediction: we remove the POI-Net and the cross-POI association, using the User-Net and the cross-user association for prediction.

Table~\ref{table: result of variants} shows the result of the ablation study.
From the table, we can observe that the variant without user prediction block gets the worst performance while the BSDA achieves the best performance.
The variant without the cross-POI association and the one without the cross-user association perform slightly worse than BSDA but better than the variant without user prediction.
The result demonstrates that each key component in BSDA plays important role in improving the location prediction performance.

\begin{figure}[t]
	\centering
	\footnotesize{
	\begin{tabular}{c}
		\includegraphics[width=0.465\textwidth]{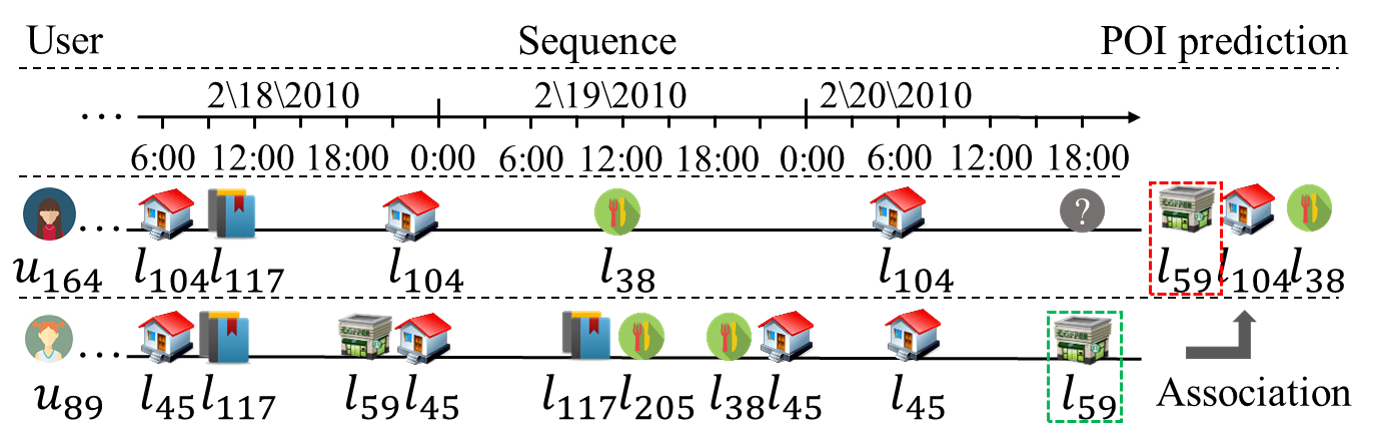} \\
		\small{(a) Case for Cross-User Association} \\	
		\includegraphics[width=0.465\textwidth]{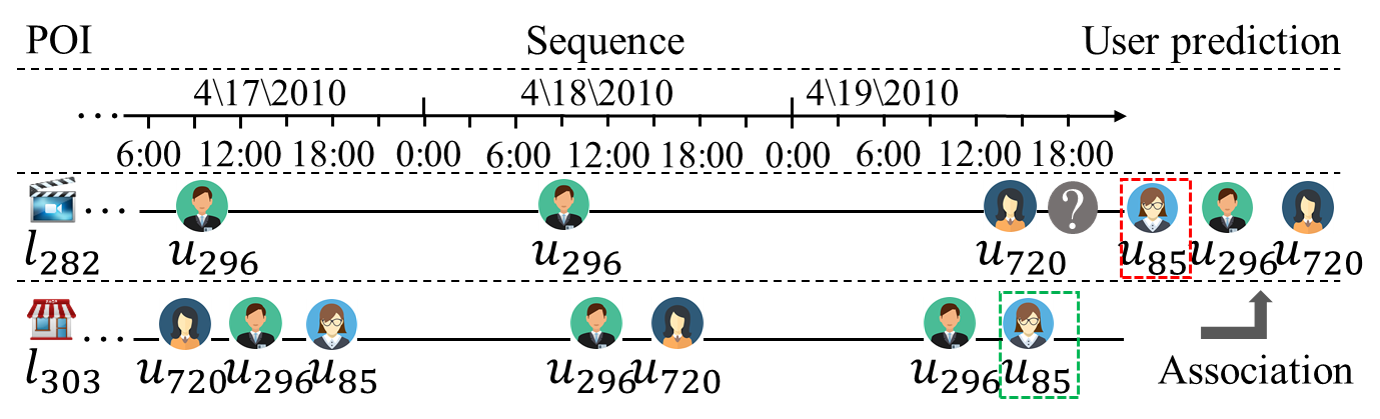} \\
		\small{(b) Case for Cross-POI Association} \\
		\includegraphics[width=0.465\textwidth]{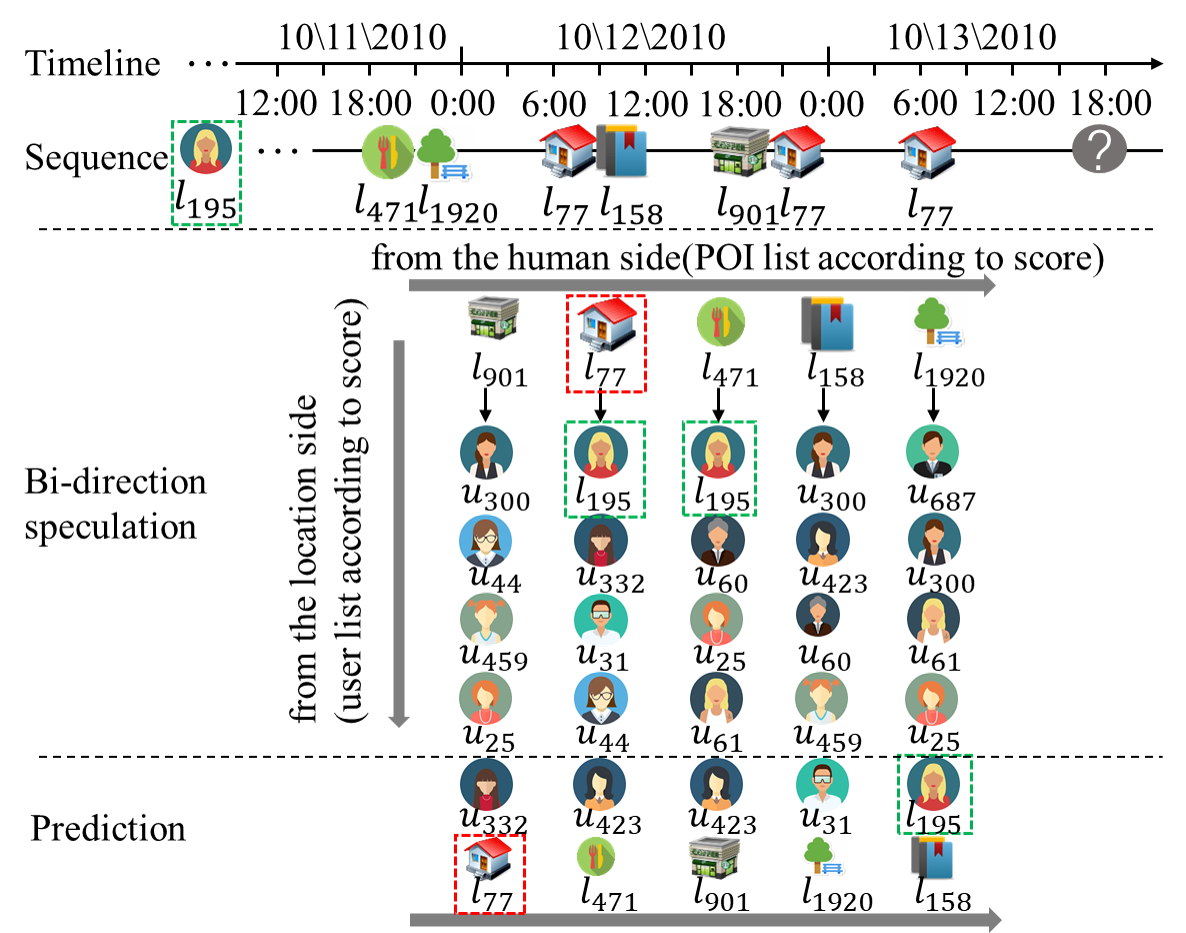} \\
        \small{(c) Case for Bi-direction Speculation} \\
	\end{tabular}}
	\caption{Prediction cases by our proposed BSDA. The ground truth is in the red dotted box.}
	\label{fig:case study}
\end{figure}

\subsubsection{Discussions}
In addition to the comparisons above, we also conduct some experiments for the model analysis further.
Due to the space limitation, we only show the results on the Gowalla dataset.

First, we evaluate the performance of our method with different fusion strategies for bi-direction speculation.
To be specific, we aggregate the results of the User-Net and POI-Net using different fusion methods including weighted addition, multiplication, maxpooling and minpooling.
Results in Table~\ref{table:fusion} demonstrate that the maxpooling is the best choice for the integration for bi-direction speculation.
It is reasonable since we intend to improve the performance of location prediction from the human side by using the user prediction from the location side. In this case, we can adjust the POI candidate score by giving a bigger score to the possible candidate or a small score to the impossible candidate.
\begin{table}[t]
\begin{center}
\small
\begin{tabular}{c|cccc}
\hline
\multirow{2}{*}{Methods}  & \multicolumn{4}{c}{Gowalla}	\\
\cline{2-5}
~  & Acc@1 & Acc@5 & Acc@10 &MRR \\
\hline
POI Prediction   &0.1215   &0.2919 &0.3721  &0.2001  \\
User Prediction  &0.1013   &0.1827  &0.2187 &0.1395    \\
\hline
Addition (0.1; 0.9)        &0.1023    &0.2789   &0.3908   &0.1811  \\
Addition (0.3; 0.7)        &0.1206    &0.3418   &0.4191  &0.2209  \\
Addition (0.5; 0.5)        & 0.1289  & 0.3578  & 0.4153  & 0.2295  \\
Addition (0.7; 0.3)        &0.1297    &0.3415   &0.4052   &0.2248  \\
Addition (0.9; 0.1)        &0.1402    &0.3291   &0.3965   &0.2364  \\
Multiplication  &0.0982   & 0.2226  & 0.3562  & 0.1693  \\
Minpooling  &0.0978  &0.1160 &0.1171   &0.4335  \\
BSDA (Maxpooling) & \textbf{0.1454} &\textbf{0.3531} &\textbf{0.4192} &\textbf{0.2413} \\
\hline
\end{tabular}
\end{center}
\caption{Results of each prediction network and methods with different fusion strategies. The numbers in parentheses following Addition denote the weight of results of POI prediction and user prediction separately.}
\label{table:fusion}
\end{table}
Besides, we evaluate the two prediction networks separately. In the BSDA, we take the raw data of the User-net for the user prediction in the POI-net. From Table~\ref{table:fusion}, we can find that the performance of the user prediction on the raw data is unsatisfactory. This is because the approach of data pre-processing is to eliminate the inactive users with few check-ins, not considering selecting POIs for the user prediction in POI-net.
Therefore, the POI-net for the user prediction will perform badly when most POIs have few check-ins.
Despite this, POI-net contributes to the performance of BSDA. 
It outputs the probability list of users who will visit a given POI at a given time.
A POI may appeal a certain group, making user candidates within a limited scope. Therefore, time-aware probability can tell users apart. When we make bi-direction speculation, collaborative analysis is effective since the results of user prediction can help find abnormal values and then adjust.

Moreover, we discuss how to measure the similarity between the two POIs.
In the literature, spatial influence is commonly considered as an essential factor in the next location prediction, where the spatial influence usually means the geographical distance.
This is easy to understand that one’s decision about where to go next is closely related to the last few visits. It may contribute more to the decision if the recent check-in is close to the current location while less if it is far away. Obviously, this spatial influence is crucial for capturing transitions between successive check-ins.
However, the purpose of the cross-POI association is to find more possible user candidates according to the most similar POIs under the assumption that similar places may appeal to common users.
So we need to find out the characteristics of POIs that appeal to common users instead of the geographical distance.

\subsection{Case Study}
To give a clear illustration of the performance of BSDA, Figure~\ref{fig:case study} shows three cases corresponding to each component of BSDA. The ground truth is in the red dotted box.
In Figure~\ref{fig:case study}(a), we display the case for the cross-user association.
Given the check-in trajectory of user $u_{164}$ , we intend to predict the next location.
But it is a challenging task since the next location is $l_{59}$ where $u_{164}$ has never checked in. In this case, we find $u_{164}$'s most similar user $u_{89}$ to deal with the problem because we believe that similar users may visit the same place.
Therefore, benefiting from the check-in history of $u_{89}$, we predict the next location $l_{59}$ successfully.

Furthermore, in Figure~\ref{fig:case study}(b), we display the case for the cross-POI association.
Similarly, we may fail to predict the next user only based on $l_{282}$'s visiting records since the data is too sparse to involve user $u_{85}$. 
In this case, $l_{303}$, as $l_{282}$'s most similar POI, provides the user candidate for us and helps to predict the next user $u_{85}$ successfully.

Finally, Figure~\ref{fig:case study}(c) illustrates how the bi-direction speculation helps the next location prediction.
Specifically, form the perspective of the human side, we give a top-5 POI candidate ranking list, where $l_{901}$ is the first while the ground truth $l_{77}$ is the second.
However, from the perspective of the location side, we infer that $u_{195}$ is unlikely to visit $l_{901}$ in the near future according to its top-5 user candidates.
Therefore, combining the results of both perspectives, we predict the next location $l_{77}$ successfully.

\section{Conclusion and Future Work}
In this paper, we propose a novel Bi-direction Speculation and Dual-level Association method (BSDA) for the next location prediction by mining both users’ interests in POIs and POIs’ appeal to users.
Specifically, we first develop two networks to explore users’ interests and POIs’ appeal separately. 
Additionally, we exploit the cross-user and cross-POI association to alleviate the data sparsity by mining similar users and POIs to enrich the candidates.
Evaluations on two public real-world datasets show that BSDA achieves significant improvements over the state-of-the-art methods.
To our knowledge, we are the first one to propose the bi-direction speculation for the location prediction, \textit{i.e.}, from both the perspective of the human side and the location side.

The function of each component of BSDA can be analyzed separately as well as the whole method, and possible refinement of each component may be future research.
In the future, we plan to extend BSDA by incorporating auxiliary information.
Besides, the performance of the next user prediction in POI-Net is unsatisfactory. Thus we will try to improve its performance for benefiting the location prediction better.
Furthermore, we will find more effective measures of user and POI similarity to provide more convincing candidates.

\section*{Acknowledgments}
This work was supported by the National Nature Science Foundation of China (U1803262).

\bibliographystyle{named}
\bibliography{ijcai21}

\end{document}